\documentclass[10pt,final,journal]{IEEEtran}
\usepackage[english]{babel}
\usepackage{mathrsfs}
\usepackage{amssymb}
\usepackage{amsmath}

\newcommand{\Aut}{\ensuremath{\mathrm{Aut}}}
\newcommand{\Sym}{\ensuremath{\mathrm{Sym}}}
\newcommand{\Ker}{\ensuremath{\mathrm{Ker}}}
\newcommand{\st}{\ensuremath{\mathrm{ST}}}
\newcommand{\supp}{\ensuremath{\mathrm{supp}}}
\newcommand{\wt}{\ensuremath{\mathrm{wt}}}
\newcommand{\F}{\ensuremath{\mathbb{F}}}
\newtheorem{Thm}{Theorem}
\newtheorem{Cor}[Thm]{Corollary}

\begin{document}
\bibliographystyle{IEEEtran}

\title{The Perfect Binary One-Error-Correcting Codes of Length $15$:
Part II---Properties}
\author{Patric R. J. {\"O}sterg{\aa}rd, Olli Pottonen, Kevin T. Phelps
\thanks{This work was supported in part by the Graduate School in
Electronics, Telecommunication and Automation, by the Nokia Foundation, and by
the Academy of Finland, Grant Numbers 107493, 110196, 
and 130142.}
\thanks{P. R. J. {\"O}sterg{\aa}rd is with the
Department of Communications and Networking, 
Aalto University, P.O.\ Box 13000, 00076 Aalto, Finland (e-mail:
patric.ostergard@tkk.fi).}
\thanks{O. Pottonen was with the
Department of Communications and Networking, Helsinki University of
Technology TKK, P.O.\ Box 3000, 02015 TKK, Finland.
He is now with the 
Finnish Defence Forces Technical Research Centre,
P.O.\ Box 10, 11311 Riihim\"aki, Finland.
(e-mail: olli.pottonen@iki.fi).}
\thanks{K. T. Phelps is
with the Department of Mathematics and Statistics, Auburn University,
Auburn, AL 36849, USA (e-mail: phelpkt@auburn.edu).}}

\maketitle
\begin{abstract}
A complete classification of the perfect binary one-error-correcting
codes of length $15$ as well as their extensions of length $16$ was
recently carried out in [P. R. J. \"Osterg\aa rd and 
O. Pottonen, ``The perfect binary one-error-correcting codes of length 
$15$: Part I---Classification,'' \emph{IEEE Trans.\ Inform.\ Theory} vol.\ 55,
pp.\ 4657--4660, 2009]. In the
current accompanying work, the classified codes are studied 
in great detail, and
their main properties are tabulated. The results include the
fact that 33 of the 80 Steiner triple systems of order 15 occur in
such codes. Further understanding is gained on full-rank codes via
switching, as it turns out that all but two full-rank codes
can be obtained through a series of such transformations from the
Hamming code. Other topics studied include (non)systematic
codes, embedded one-error-correcting codes, 
and defining sets of codes. A classification of certain mixed 
perfect codes is also obtained.
\end{abstract}

\begin{IEEEkeywords}
classification, Hamming code, perfect code, 
Steiner system, switching
\end{IEEEkeywords}

\section{Introduction}

\IEEEPARstart{W}{e} consider binary codes of length $n$ over the Galois field $\F_2$, that is,
subsets $C \subseteq \F^n_2$. The \emph{(Hamming) distance} 
$d({\bf x}, {\bf y})$ between two words
${\bf x}$, ${\bf y}$ is the number of coordinates in which they differ,
and the \emph{(Hamming) weight} $\wt({\bf x})$ 
of a word ${\bf x}$
is the number of nonzero
coordinates. The \emph{support} of a word is the set of nonzero
coordinates, that is, $\supp({\bf x}) = \mbox{$\{i : x_i \neq 0\}$}$. Accordingly,
$d({\bf x}, {\bf y}) = \wt({\bf x}-{\bf y}) = |\supp({\bf x}-{\bf y})|$.

The \emph{minimum distance} of a code is the largest integer $d$
such that the distance between any distinct codewords
is at least $d$. The balls of radius $\lfloor (d-1)/2 \rfloor$
centered around the codewords of a code with minimum distance $d$ 
are nonintersecting, so such a code is said to be 
a $\lfloor (d-1)/2 \rfloor$-error-correcting code. If these balls
simultaneously pack and cover the ambient space, then the code is
called \emph{perfect}. A $t$-error-correcting perfect code is also
called a \emph{$t$-perfect code}.

It is well known \cite{MS77} that binary perfect codes 
exist exactly for $d=1$; $d=n$; $d=(n-1)/2$ for odd $n$;
$d=3, n=2^m-1$ for $m\ge 2$; and $d=7, n=23$. The first three types of codes
are called trivial, the fourth has the parameters of Hamming codes,
and the last one is the binary Golay code. 

The number of binary 1-perfect codes of length 15 was recently determined in
\cite{OP09} (where the codes are also made available
in electronic form) using a constructive approach. It turned out that there are
$5\,983$ inequivalent such codes, and these have $2\,165$ inequivalent
extensions. Two binary codes are said to be
\emph{equivalent} if one can be obtained from the other by permuting coordinates
and adding a constant vector. Such a mapping that produces a code from
itself is an \emph{automorphism}; the set of all automorphisms of a code
form a group, the \emph{automorphism group}.

The complete set of inequivalent codes is a valuable tool that
makes it possible to study a wide variety of properties. Our aim is
to answer questions stated in \cite{EV98,H08} and elsewhere, and
in general to gain as good understanding as possible of the properties of 
the binary 1-perfect codes of length 15. The graph isomorphism program
\emph{nauty} \cite{M90} played a central role in several of the
computations.

For completeness, we give the table with the distribution of
automorphism group orders
from \cite{OP09} in Section~\ref{sect:auto}, where also  
the distribution of kernels is tabulated (including some corrections
to earlier results).
The supports of the differences between a codeword in a 
1-perfect code and other codewords at (minimum) 
distance 3 form a Steiner triple system.
In Section~\ref{sect:steiner}, such occurrences of Steiner
triple systems in the codes---and occurrences of Steiner quadruple systems in the extended
codes---are studied, determining among other things that exactly 33 of
the 80 Steiner triple systems of order 15 occur in these 1-perfect codes. 
Other topics addressed include the determination of
the largest number of isomorphism classes of 
Steiner triple systems in a code.

In Section~\ref{sect:i} partial results are provided 
on perhaps the most intriguing issue
regarding 1-perfect codes, namely that of finding constructions 
(explanations) for all different codes. It turns out that
the binary 1-perfect codes of length 15 are partitioned into
just 9 switching classes. The technique of switching 
is utilized also in Section~\ref{sect:define}, 
for proving general results for defining sets of 1-perfect codes.
(Non)systematic \mbox{1-perfect}
codes are treated in Section~\ref{sect:syst}, and 
embedded one-error-correcting codes 
and related orthogonal arrays are considered
in Section~\ref{sect:embed}.

Many classes of mixed perfect codes with alphabet sizes that 
are powers of 2 are classified in Section~\ref{sect:mixed}.
The paper is concluded in Section~\ref{sect:other}, which
includes a list of a few interesting problems related to binary
1-perfect codes of length 15, yet unanswered.

\section{Automorphism Groups}

\label{sect:auto}

First, we give formal definitions of several central concepts,
some of which were briefly mentioned in the Introduction.
A permutation $\pi$ of the set $\{1,2,\ldots,n\}$ acts on codewords
by permuting the coordinates in the obvious manner.
Pairs $(\pi, {\bf x})$ form the
\emph{wreath product} $S_2\wr S_n$, which acts on codes as
$(\pi, {\bf x})(C) = \pi(C + {\bf x}) = \pi(C) + \pi({\bf x})$.
Two codes, $C_1$ and $C_2$, are said to be 
\emph{isomorphic} if $C_1 = \pi(C_2)$ for some $\pi$ and 
\emph{equivalent}
if $C_1 = \pi(C_2 + {\bf x})$ for some $\pi, {\bf x}$.

The \emph{automorphism group} of a code $C$, $\Aut(C)$, is the group
of all pairs $(\pi, {\bf x})$ such that $C = \pi(C + {\bf x})$.
Two important subgroups of $\Aut(C)$ are the \emph{group of symmetries},
\[
\Sym(C) = \{\pi : \pi(C) = C\}
\]
and the \emph{kernel}
\[
\Ker(C) = \{{\bf x} : C + {\bf x} = C\}.
\]
If the code contains the all-zero word, ${\bf 0}$, then the elements of
the kernel are codewords. The distribution of the orders of
the automorphism groups of the binary 1-perfect codes of length
15 and their extensions are presented in Table~\ref{tbl:perfautorder}
and Table~\ref{tbl:extautorder}, respectively.

\begin{table}[htbp]
\begin{center}
\caption{Automorphism groups of codes}
\label{tbl:perfautorder}
\begin{tabular}{rrrrrr}\hline
$|\Aut(C)|$ & \# & $|\Aut(C)|$ & \# & $|\Aut(C)|$ & \# \\ \hline
8   &      3 &     512 & 1\,017 &      24\,576 & 7  \\
12  &      3 &     672 &      3 &      32\,768 & 8  \\
16  &      5 &     768 &     32 &      43\,008 & 4  \\
24  &     10 &  1\,024 &    697 &      49\,152 & 10 \\
32  &    138 &  1\,536 &     17 &      65\,536 & 5  \\
42  &      2 &  2\,048 &    406 &      98\,304 & 1  \\
48  &     12 &  2\,688 &      1 &     131\,072 & 1  \\
64  &    542 &  3\,072 &     37 &     172\,032 & 1  \\
96  &     22 &  3\,840 &      1 &     196\,608 & 5  \\
120 &      1 &  4\,096 &    202 &     344\,064 & 2  \\
128 & 1\,230 &  5\,376 &      4 &     393\,216 & 2  \\
192 &     18 &  6\,144 &     35 &     589\,824 & 1  \\
256 & 1\,319 &  8\,192 &     94 & 41\,287\,680 & 1  \\
336 &      3 & 12\,288 &      7 \\
384 &     30 & 16\,384 &     44 \\
\hline
\end{tabular}
\end{center}
\end{table}

\begin{table}[htbp]
\begin{center}
\caption{Automorphism groups of extended codes}
\label{tbl:extautorder}
\begin{tabular}{rrrrrr}\hline
$|\Aut(C)|$ & \# & $|\Aut(C)|$ & \# & $|\Aut(C)|$ & \# \\ \hline
128    &  11 & 5\,376   &   1 & 196\,608      & 6 \\
192    &   5 & 6\,144   &  23 & 262\,144      & 3\\
256    & 105 & 8\,192   & 174 & 344\,064      & 1\\
384    &   9 & 10\,752  &   2 & 393\,216      & 3\\
512    & 377 & 12\,288  &  22 & 524\,288      & 2\\
672    &   2 & 16\,384  & 103 & 688\,128      & 1\\
768    &  19 & 24\,576  &  12 & 786\,432      & 2\\
1\,024 & 416 & 32\,768  &  47 & 1\,572\,864   & 3\\
1\,344 &   1 & 43\,008  &   2 & 2\,359\,296   & 1\\
1\,536 &  21 & 49\,152  &  18 & 2\,752\,512   & 1\\
1\,920 &   1 & 61\,440  &   1 & 3\,145\,728   & 1\\
2\,048 & 394 & 65\,536  &  33 & 5\,505\,024   & 2\\
2\,688 &   1 & 86\,016  &   3 & 6\,291\,456   & 1\\
3\,072 &  18 & 98\,304  &  12 & 660\,602\,880 & 1\\
4\,096 & 298 & 131\,072 &   6 \\
\hline
\end{tabular}
\end{center}
\end{table}

The orbits of codewords of the binary 1-perfect codes 
of length 15 and their extensions are
tabulated in Tables~\ref{tbl:prforbits} and \ref{tbl:extorbits}, respectively.
Here the notation $g_1^{a_1}g_2^{a_2}\cdots g_m^{a_m}$ means that the number
of orbits of size $g_i$ is $a_i$.

\begin{table*}[htbp]
\begin{center}
\caption{Orbits of perfect codes}
\label{tbl:prforbits}
\begin{tabular}{rrrrrrrrrr}\hline
Orbits & \# & Orbits & \# & Orbits & \# & Orbits & \# & Orbits & \# \\\hline
$8^{256}$ & 3 & $32^{16}64^{24}$ & 30 & $64^{8}192^{8}$ & 6 & $128^{4}256^{4}512^{1}$ & 4 & $128^{2}896^{2}$ & 2\\
$4^{8}12^{168}$ & 3 & $64^{32}$ & 539 & $32^{4}96^{4}192^{8}$ & 5 & $256^{6}512^{1}$ & 64 & $128^{8}1024^{1}$ & 1\\
$16^{128}$ & 65 & $32^{4}48^{32}96^{4}$ & 3 & $64^{2}192^{10}$ & 5 & $64^{16}512^{2}$ & 4 & $128^{4}256^{2}1024^{1}$ & 1\\
$8^{16}24^{80}$ & 2 & $16^{8}48^{8}96^{16}$ & 2 & $64^{8}128^{8}256^{2}$ & 25 & $64^{8}128^{4}512^{2}$ & 4 & $256^{4}1024^{1}$ & 14\\
$8^{4}24^{84}$ & 8 & $32^{4}96^{20}$ & 20 & $128^{12}256^{2}$ & 215 & $128^{8}512^{2}$ & 25 & $256^{2}512^{1}1024^{1}$ & 22\\
$16^{96}32^{16}$ & 43 & $24^{2}40^{8}120^{14}$ & 1 & $64^{16}256^{4}$ & 7 & $128^{4}256^{2}512^{2}$ & 23 & $512^{2}1024^{1}$ & 35\\
$16^{64}32^{32}$ & 83 & $32^{16}64^{16}128^{4}$ & 9 & $64^{8}128^{4}256^{4}$ & 2 & $256^{4}512^{2}$ & 339 & $256^{1}768^{1}1024^{1}$ & 3\\
$16^{32}32^{48}$ & 5 & $64^{24}128^{4}$ & 301 & $128^{8}256^{4}$ & 573 & $128^{2}384^{2}512^{2}$ & 3 & $1024^{2}$ & 218\\
$32^{64}$ & 249 & $32^{32}128^{8}$ & 5 & $128^{4}256^{6}$ & 117 & $256^{2}512^{3}$ & 111 & $768^{1}1280^{1}$ & 1\\
$2^{2}14^{14}42^{44}$ & 2 & $32^{16}64^{8}128^{8}$ & 1 & $256^{8}$ & 406 & $512^{4}$ & 223 & $256^{2}1536^{1}$ & 8\\
$16^{8}48^{40}$ & 12 & $64^{16}128^{8}$ & 624 & $16^{2}112^{6}336^{4}$ & 3 & $32^{1}224^{3}672^{2}$ & 3 & $512^{1}1536^{1}$ & 11\\
$16^{2}48^{42}$ & 1 & $64^{8}128^{12}$ & 101 & $64^{8}384^{4}$ & 1 & $128^{4}768^{2}$ & 4 & $256^{1}1792^{1}$ & 7\\
$32^{48}64^{8}$ & 147 & $128^{16}$ & 626 & $128^{4}384^{4}$ & 44 & $256^{2}768^{2}$ & 40 & $2048^{1}$ & 201\\
$32^{32}64^{16}$ & 292 & $64^{2}96^{16}192^{2}$ & 12 & $256^{2}384^{4}$ & 4 & $512^{1}768^{2}$ & 10 & \\
\hline
\end{tabular}
\end{center}
\end{table*}

\begin{table*}[htbp]
\begin{center}
\caption{Orbits of extended codes}
\label{tbl:extorbits}
\begin{tabular}{rrrrrrrr}\hline
Orbits & \# & Orbits & \# & Orbits & \# & Orbits & \# \\\hline
$32^{32}64^{16}$ & 2 & $128^{10}256^{3}$ & 6 & $64^{2}128^{5}256^{3}512^{1}$ & 6 & $384^{2}512^{1}768^{1}$ & 6\\
$64^{32}$ & 21 & $64^{16}256^{4}$ & 2 & $128^{6}256^{3}512^{1}$ & 8 & $128^{4}768^{2}$ & 2\\
$64^{24}128^{4}$ & 40 & $64^{8}128^{4}256^{4}$ & 27 & $64^{4}128^{2}256^{4}512^{1}$ & 1 & $128^{2}256^{1}768^{2}$ & 3\\
$64^{16}128^{8}$ & 75 & $64^{4}128^{6}256^{4}$ & 7 & $64^{2}128^{3}256^{4}512^{1}$ & 6 & $256^{2}768^{2}$ & 16\\
$64^{14}128^{9}$ & 1 & $64^{2}128^{7}256^{4}$ & 2 & $128^{4}256^{4}512^{1}$ & 80 & $512^{1}768^{2}$ & 7\\
$64^{12}128^{10}$ & 9 & $128^{8}256^{4}$ & 121 & $64^{2}128^{1}256^{5}512^{1}$ & 2 & $128^{2}896^{2}$ & 1\\
$16^{2}32^{3}64^{6}128^{12}$ & 1 & $64^{4}128^{4}256^{5}$ & 13 & $128^{2}256^{5}512^{1}$ & 10 & $64^{8}128^{4}1024^{1}$ & 1\\
$64^{8}128^{12}$ & 13 & $64^{2}128^{5}256^{5}$ & 6 & $256^{6}512^{1}$ & 53 & $64^{4}128^{6}1024^{1}$ & 1\\
$16^{2}32^{3}64^{4}128^{13}$ & 1 & $128^{6}256^{5}$ & 20 & $128^{8}512^{2}$ & 14 & $128^{6}256^{1}1024^{1}$ & 1\\
$64^{6}128^{13}$ & 6 & $64^{8}256^{6}$ & 8 & $64^{4}128^{4}256^{1}512^{2}$ & 1 & $64^{2}128^{3}256^{2}1024^{1}$ & 1\\
$64^{4}128^{14}$ & 2 & $64^{4}128^{2}256^{6}$ & 1 & $64^{2}128^{5}256^{1}512^{2}$ & 4 & $128^{4}256^{2}1024^{1}$ & 7\\
$64^{2}128^{15}$ & 2 & $64^{2}128^{3}256^{6}$ & 2 & $64^{2}128^{3}256^{2}512^{2}$ & 2 & $64^{2}128^{1}256^{3}1024^{1}$ & 2\\
$128^{16}$ & 55 & $128^{4}256^{6}$ & 93 & $128^{4}256^{2}512^{2}$ & 82 & $128^{2}256^{3}1024^{1}$ & 4\\
$64^{2}96^{16}192^{2}$ & 2 & $64^{2}128^{1}256^{7}$ & 5 & $64^{2}128^{1}256^{3}512^{2}$ & 3 & $256^{4}1024^{1}$ & 49\\
$32^{2}64^{1}96^{10}192^{5}$ & 1 & $128^{2}256^{7}$ & 6 & $128^{2}256^{3}512^{2}$ & 21 & $128^{2}384^{2}1024^{1}$ & 3\\
$32^{4}96^{8}192^{6}$ & 1 & $256^{8}$ & 75 & $256^{4}512^{2}$ & 126 & $128^{4}512^{1}1024^{1}$ & 2\\
$48^{2}64^{2}96^{5}192^{7}$ & 1 & $64^{2}96^{4}192^{6}384^{1}$ & 2 & $128^{4}512^{3}$ & 27 & $128^{2}256^{1}512^{1}1024^{1}$ & 14\\
$32^{4}96^{4}192^{8}$ & 1 & $16^{1}48^{1}64^{3}96^{2}128^{2}192^{1}384^{3}$ & 1 & $128^{2}256^{1}512^{3}$ & 10 & $256^{2}512^{1}1024^{1}$ & 33\\
$64^{2}96^{4}192^{8}$ & 2 & $96^{2}128^{1}192^{3}384^{3}$ & 1 & $256^{2}512^{3}$ & 54 & $512^{2}1024^{1}$ & 50\\
$64^{20}128^{4}256^{1}$ & 4 & $96^{2}128^{2}192^{1}256^{1}384^{3}$ & 1 & $128^{1}384^{1}512^{3}$ & 1 & $128^{2}768^{1}1024^{1}$ & 4\\
$64^{12}128^{8}256^{1}$ & 63 & $16^{1}48^{1}64^{3}128^{2}384^{4}$ & 1 & $512^{4}$ & 111 & $256^{1}768^{1}1024^{1}$ & 5\\
$64^{4}128^{12}256^{1}$ & 16 & $128^{4}384^{4}$ & 4 & $320^{2}384^{2}640^{1}$ & 1 & $1024^{2}$ & 81\\
$128^{14}256^{1}$ & 4 & $96^{2}128^{1}192^{1}384^{4}$ & 1 & $32^{1}224^{3}336^{2}672^{1}$ & 2 & $768^{1}1280^{1}$ & 1\\
$64^{16}128^{4}256^{2}$ & 18 & $128^{1}192^{2}384^{4}$ & 1 & $32^{1}224^{1}448^{1}672^{2}$ & 1 & $128^{4}1536^{1}$ & 1\\
$64^{8}128^{8}256^{2}$ & 54 & $256^{2}384^{4}$ & 6 & $32^{2}96^{2}128^{2}192^{4}768^{1}$ & 1 & $128^{2}256^{1}1536^{1}$ & 8\\
$64^{4}128^{10}256^{2}$ & 6 & $64^{4}128^{8}256^{1}512^{1}$ & 2 & $32^{1}64^{1}96^{1}128^{1}192^{5}768^{1}$ & 1 & $256^{2}1536^{1}$ & 12\\
$128^{12}256^{2}$ & 78 & $128^{10}256^{1}512^{1}$ & 2 & $32^{1}96^{1}128^{1}192^{4}256^{1}768^{1}$ & 1 & $512^{1}1536^{1}$ & 10\\
$64^{12}128^{4}256^{3}$ & 26 & $64^{8}128^{4}256^{2}512^{1}$ & 3 & $32^{1}96^{1}128^{3}384^{2}768^{1}$ & 1 & $16^{1}112^{1}128^{1}1792^{1}$ & 1\\
$64^{8}128^{6}256^{3}$ & 20 & $64^{4}128^{6}256^{2}512^{1}$ & 8 & $128^{4}384^{2}768^{1}$ & 1 & $128^{2}1792^{1}$ & 2\\
$64^{6}128^{7}256^{3}$ & 5 & $128^{8}256^{2}512^{1}$ & 58 & $128^{2}256^{1}384^{2}768^{1}$ & 21 & $32^{1}224^{1}1792^{1}$ & 1\\
$64^{4}128^{8}256^{3}$ & 33 & $64^{8}128^{2}256^{3}512^{1}$ & 4 & $64^{1}192^{1}256^{1}384^{2}768^{1}$ & 1 & $256^{1}1792^{1}$ & 5\\
$64^{2}128^{9}256^{3}$ & 8 & $64^{4}128^{4}256^{3}512^{1}$ & 12 & $256^{2}384^{2}768^{1}$ & 5 & $2048^{1}$ & 101\\
\hline
\end{tabular}
\end{center}
\end{table*}

It has been known since the early days of coding theory 
\cite{L57,SS59} that binary 1-perfect codes
are \emph{distance invariant}, that is, the distance distribution
of the other codewords with respect to any codeword does not
depend on the choice of codeword. In particular,
there is always one codeword at distance $n$, that is,
the all-one word is in the kernel of
all binary 1-perfect codes; the codes
are said to be \emph{self-complementary}. The distance distribution
for binary 1-perfect codes of length 15 is
\[
1\ 0\ 0\ 35\ 105\ 168\ 280\ 435\ 435\ 280\ 168\ 105\ 35\ 0\ 0\ 1.
\]
There is also only one distance distribution with respect to any word
that is not a codeword of such a code:
\[
0\ 1\ 7\ 28\ 84\ 189\ 315\ 400\ 400\ 315\ 189\ 84\ 28\ 7\ 1\ 0.
\]
 
Once the equivalence classes of codes have been classified,
classifying the isomorphism classes is straightforward.
Isomorphic codes necessarily belong to the same 
equivalence class, so representatives from the
isomorphism classes can be obtained by
translating representatives from the equivalence classes.
The following theorem characterizes the situation further.

\begin{Thm}
The codes $C + {\bf x}$ and $C + {\bf y}$ are isomorphic if and only if
${\bf x}$ and ${\bf y}$ are in the same $\Aut(C)$-orbit.
\end{Thm}
\begin{IEEEproof}
  The codes are isomorphic iff there is a permutation $\pi$ such that
  $\pi(C + {\bf x}) = C + {\bf y}$, which is equivalent to $C = \pi(C
  + {\bf x} + \pi^{-1}({\bf y}))$. The last equation holds iff $(\pi,
  {\bf x} + \pi^{-1}({\bf y})) \in \Aut(C)$. Clearly this pair maps
  ${\bf x}$ to ${\bf y}$. Conversely, every pair which maps ${\bf x}$
  to ${\bf y}$ is of the type $(\pi, {\bf x} + \pi^{-1}({\bf y}))$
  with $\pi$ arbitrary.
\end{IEEEproof}

There are $1\,637\,690$ isomorphism classes of binary
1-perfect codes of length 15, $139\,350$ of
which contain the all-zero codeword. The groups of symmetries 
of these codes
are tabulated in Table~\ref{tbl:perfsymorder}. The extended 
1-perfect
codes have $347\,549$ isomorphism classes, of which $22\,498$,
$139\,350$, and $185\,701$ contain a codeword with minimum weight $0$, $1$,
and $2$, respectively. The groups of symmetries of these codes are listed in
Table~\ref{tbl:extsymorder}.

\begin{Thm}\label{Thm:symfixed}
  An (extended) binary 1-perfect code $C$ contains an embedded (extended) binary
  1-perfect code on the coordinates that are fixed by any subgroup
  $G \subseteq \Sym(C)$.
\end{Thm}

\begin{IEEEproof}
  Let $C$ be a binary 1-perfect code and $T$ the set of coordinates
  not fixed by $G$, and let $H$ be the set of all words that
  have zeros for all coordinates in $T$. Now consider the
  embedded code $C' = C \cap H$. The code $C'$ 
  is 1-perfect (after deleting the coordinates in $T$) if
  every word in $H$ is at distance at most $1$ from a codeword 
  in $C'$. Now assume that
  this is not the case, that is, that there is a word 
  ${\bf x} \in H$ that is at distance at least 2 from
  all words in $C'$. 

  Since $C$ is a 1-perfect code, there 
  must be a codeword ${\bf y} \in C \setminus C'$ such
  that $d({\bf x}, {\bf y}) = 1$. Moreover, since 
  ${\bf y} \not\in C'$, it follows that 
  $|\supp({\bf y}) \cap T| = 1$. As there is a $\pi \in
  G$ such that $\pi({\bf y}) \neq {\bf y}$ and $\pi$ 
  preserves the weight within $T$, we get that
  $d({\bf y}, \pi({\bf y})) = 2$. This is a contradiction
  since both ${\bf y}$ and $\pi({\bf y})$ are codewords and
  $C$ has minimum distance 3.

  To prove the claim for an extended binary
  1-perfect code $C$, first
  puncture the code at any coordinate fixed by $G$ and use the 
  previous result for binary 1-perfect codes; 
  extension of the embedded 1-perfect 
  code thereby obtained indeed gives a subcode of $C$ (as all
  coordinates that are deleted have value 0 for this subcode).
\end{IEEEproof}

Note that Theorem~\ref{Thm:symfixed} can be generalized 
by instead of $\Sym(C)$ considering the subgroup of
$\Aut(C + {\bf x})$ that stabilizes ${\bf x}$ for any word 
${\bf x}$.  Also note that Theorem~\ref{Thm:symfixed} implies 
that $\Sym(C)$ has $2^k-1$ fixed coordinates for any 
binary \mbox{1-perfect} 
code $C$, and $0$ or $2^k$ fixed coordinates for any extended 
binary 1-perfect code $C$. The numbers of fixed coordinates are 
tabulated in Tables~\ref{tbl:perffixed} and \ref{tbl:extfixed}.

\begin{table}[htbp]
\begin{center}
\caption{Groups of symmetries of  codes}
\label{tbl:perfsymorder}
\begin{tabular}{rrrrrr}\hline
$|\Sym(C)|$ & \# & $|\Sym(C)|$ & \# & $|\Sym(C)|$ & \# \\ \hline
1  & 668\,929 & 12  &     80 & 96      & 37 \\
2  & 646\,808 & 16  & 2\,222 & 168     &  3 \\
3  &   2\,598 & 21  &     45 & 192     & 32 \\
4  & 288\,221 & 24  &    536 & 288     &  1 \\
5  &        3 & 32  &    685 & 1\,344  &  7 \\
6  &       64 & 48  &     24 & 20\,160 &  1 \\
8  &  27\,370 & 64  &     24 & \\
\hline
\end{tabular}
\end{center}
\end{table}

\begin{table}[htbp]
\begin{center}
\caption{Groups of symmetries of extended codes}
\label{tbl:extsymorder}
\begin{tabular}{rrrrrr}\hline
$|\Sym(C)|$ & \# & $|\Sym(C)|$ & \# & $|\Sym(C)|$ & \# \\ \hline
 1 &   43\,935 &  42 &      8 &      512 & 25 \\
 2 &  111\,372 &  48 &    224 &      768 & 17 \\
 3 &       768 &  64 & 1\,012 &   1\,152 & 1  \\
 4 &   98\,199 &  80 &      1 &   1\,344 & 3  \\
 5 &         5 &  96 &    137 &   1\,536 & 10 \\
 6 &       613 & 128 &    394 &   2\,688 & 1  \\
 8 &   57\,502 & 168 &      6 &   3\,072 & 8  \\
12 &       390 & 192 &     44 &  20\,160 & 1  \\
16 &   25\,858 & 256 &     80 &  21\,504 & 7  \\
21 &        30 & 288 &      1 & 322\,560 & 1  \\
24 &       307 & 336 &     15 \\
32 &    6\,508 & 384 &     66 \\
\hline
\end{tabular}
\end{center}
\end{table}

\begin{table}[htbp]
\begin{center}
\caption{Coordinates fixed by symmetries of codes}
\label{tbl:perffixed}
\begin{tabular}{rrrrrr}\hline
Coordinates & \# & Coordinates & \# & Coordinates & \# \\ \hline
0 &  13 & 3 &  37\,732 & 15 & 668\,929 \\
1 & 818 & 7 & 930\,198 & \\
\hline
\end{tabular}
\end{center}
\end{table}

\begin{table}[htbp]
\begin{center}
\caption{Coordinates fixed by symmetries of extended codes}
\label{tbl:extfixed}
\begin{tabular}{rrrrrr}\hline
Coordinates & \# & Coordinates & \# & Coordinates & \# \\ \hline
0 & 162\,499 & 2 &    519 &  8 & 131\,187 \\
1 &       17 & 4 & 9\,392 & 16 &   43\,935 \\\hline
\end{tabular}
\end{center}
\end{table}

The existence problem for binary 1-perfect codes with
automorphism group of (minimum) order 2 has received some
attention. By Table~\ref{tbl:perfautorder}, there are no such 
codes of length 15. This contradicts claims in \cite[p. 242]{H08}
regarding existence of such codes. Existence for admissible lengths
at least $2^8-1$ and an interval of ranks has been proved in 
\cite{H05}. For lengths $2^m-1$ with $m=5,6,7$, only an
8-line outline of proof has been published \cite{M98}; there
is an obvious desire for a detailed treatment of those cases.

In Table~\ref{tab:rank},
we display the number of codes with respect
to their rank and kernel size. The results for rank 15 are
new, and several entries for rank 13 and 14 correct
earlier results from \cite{ZZ04,ZZ06}
(also surveyed in \cite[p.\ 237]{H08}); the authors 
of the original papers have rechecked their results
for rank at most 14 and have arrived at results that 
corroborate those presented here.

\begin{table}[htbp]
\begin{center}
\caption{Codes by rank and kernel size}
\label{tab:rank}
\begin{tabular}{rrrrrr}\hline
Kernel$\backslash$Rank & $11$   & $12$   & $13$   & $14$   & $15$ \\\hline
$2$     &      &     &      &      & 19   \\      
$4$     &      &     &	    &163   & 14   \\
$8$     &      &     &	    &1\,287& 8    \\
$16$    &      &     & 224  &2\,334& 338  \\
$32$    &      &     & 262  &941   & 19   \\
$64$    &      &     & 176  &129   &      \\
$128$   &      & 12  & 28   &8	   &      \\
$256$   &      & 3   & 13   &1	   &      \\
$512$   &      & 3   &	    &	   &      \\
$1\,024$  &      &     &	    &	   &      \\
$2\,048$  & 1    &     &	    &	   &      \\\hline
\end{tabular}
\end{center}
\end{table}  

As can be seen, there are 398 codes with full rank.
Partial results for rank 15 can be found
in \cite{M06}. All possible kernels (unique for sizes
2, 4, and 8; two for sizes 16 and 32) of the full-rank codes 
are, up to isomorphism, generated by the words in Table~\ref{tab:kernels}.

\begin{table}
\begin{center}
\caption{Bases of kernels of full-rank codes}
\label{tab:kernels}
\begin{tabular}{lll}\hline
111111110000000 &  111111001100000 &  111100001111000 \\
110000000000100 &  000000001111111; \\[4pt]
111111100000000 &  111100011110000 &  110010011001100 \\
101011010101010 &  001110011000011; \\[4pt]
111100000000000 &  000011110000000 &  000000001111000 \\
000000000000111; \\[4pt]
111111100000000 &  111000011110000 &  100110011001100 \\
011110011000011; \\[4pt]
111111100000000 &  111000011110000 &  111000000001111; \\[4pt]
111111100000000 &  000000011111111; \\[4pt]
111111111111111.\\\hline
\end{tabular}
\end{center}
\end{table}

A \emph{tiling} of $\F_2^n$ is a pair $(V,A)$ of subsets of 
$\F_2^n$ such that every ${\bf x} \in \F_2^n$ can be written 
in exactly one way as ${\bf x} = {\bf v} + {\bf a}$ with 
${\bf v} \in V$ and ${\bf a} \in A$.
A tiling $(V,A)$ of $\F_2^n$ is said to be \emph{full rank} if
$\mbox{rank}(V)=\mbox{rank}(A)=n$ and ${\bf 0} \in V \cap A$.
The results
of the current work provide data for a classification of 
full-rank tilings of $\F_2^i$, $10 \leq i \leq 15$, where
one of the sets has size 16, cf.\ \cite{CLVZ96,EV94,EV98}.
Nonexistence of full-rank binary 1-perfect codes of length 15
with a kernel of size 64 corroborates the result in \cite{OV04} 
that there are no full-rank tilings of $\F_2^9$. Moreover,
the observation regarding the structure of full-rank tilings of 
$\F_2^{10}$ with $|V|=2^4$ and $|A|=2^6$ at the very end of
\cite{OV04} now gets an independent verification.

The number of extended binary 1-perfect codes of length 16
with respect to their rank and kernel size is shown in 
Table~\ref{tab:erank}. 

\begin{table}[htbp]
\begin{center}
\caption{Extended codes by rank and kernel size}
\label{tab:erank}
\begin{tabular}{rrrrrr}\hline
Kernel$\backslash$Rank & $11$   & $12$   & $13$   & $14$   & $15$ \\\hline
$2$     &      &     &      &      & 18   \\      
$4$     &      &     &	    & 102  & 14   \\
$8$     &      &     &	    & 449  &  8   \\
$16$    &      &     &  82  & 786  & 123  \\
$32$    &      &     &  89  & 326  & 12   \\
$64$    &      &     &  67  & 53   &      \\
$128$   &      &  8  &  11  & 4	   &      \\
$256$   &      &  2  &   7  & 1	   &      \\
$512$   &      &  2  &	    &	   &      \\
$1\,024$  &      &     &	    &	   &      \\
$2\,048$  &  1   &     &	    &	   &      \\\hline
\end{tabular}
\end{center}
\end{table}  

The kernels for the extended binary 1-perfect codes of length
16 and rank 15 are exactly those obtained by extending the
kernels for the full-rank binary 1-perfect codes of length 15,
listed in Table~\ref{tab:kernels}.

\section{Steiner Systems in 1-Perfect Codes}

\label{sect:steiner}

A \emph{Steiner system} $S(t,k,v)$ is a collection of
$k$-subsets (called \emph{blocks}) of a $v$-set of points,
such that every $t$-subset of the $v$-set is contained in
exactly one block. Steiner systems $S(2,3,v)$ and $S(3,4,v)$ are called
\emph{Steiner triple systems} and \emph{Steiner quadruple systems},
respectively, and are often referred to as STS$(v)$ and SQS$(v)$,
where $v$ is called the \emph{order} of the system.
These are related to binary 1-perfect codes in the following way.

If $C$ is a binary 1-perfect code of length $v$ and ${\bf x} \in C$, 
then the codewords of $C + {\bf x}$ with weight $3$ form a Steiner triple 
system of order $v$. Analogously, if $C$ is an extended binary 1-perfect 
code and ${\bf x} \in C$, then the codewords of $C + {\bf x}$ with weight 
$4$ form a Steiner quadruple system.

There are 80 Steiner triple systems of order 15. The longstanding open
question whether all Steiner triple systems of order $2^m-1$
occur in some binary 1-perfect code of length $2^m-1$ 
was settled in
\cite{OP07}, by showing that at least two of the 80 STS(15) do not
occur in a 1-perfect code. We are now able to determine exactly which
STS(15) occur in a binary 1-perfect code---the total number of 
such STS(15) is 33---and furthermore in how many codes
each such system occurs. This information is given 
in Table \ref{tab:sts} 
using the numbering of the STS(15) from \cite{MPR83}. 
As far as the authors are aware, existence results for all of 
these, except those with indices 25 and 26, can be found in the 
literature \cite{L95,N99,P83}.

\begin{table}[htbp]
\begin{center}
\caption{Occurrences of Steiner triple systems}
\label{tab:sts}
\begin{tabular}{rrrrrr}\hline
Index & \# & Index & \# & Index & \# \\ \hline
1 & 205    & 12 & 640   & 24 & 44  \\
2 & 1\,543 & 13 & 1\,666& 25 & 158 \\
3 & 1\,665 & 14 & 1\,268& 26 & 158 \\
4 & 3\,623 & 15 & 1\,961& 29 & 187 \\
5 & 2\,209 & 16 & 745  	& 33 & 37  \\
6 & 1\,229 & 17 & 781  	& 35 & 2   \\
7 & 335	   & 18 & 1\,653& 39 & 2   \\
8 & 3\,290 & 19 & 204  	& 54 & 2   \\
9 & 2\,950 & 20 & 493  	& 61 & 57  \\
10 & 2\,914& 21 & 50   	& 64 & 29  \\
11 & 636   & 22 & 55   	& 76 & 6   \\\hline
\end{tabular}
\end{center}
\end{table}

It is not difficult to see that all Steiner triple systems 
in a linear code are necessarily equal, so Hamming codes
show that the problem of minimizing the number of different
Steiner triple systems in a binary 1-perfect code has an obvious
solution (for all lengths $2^m-1$). On the other hand,
Table \ref{tab:spec} shows that 14 is the maximum number of
isomorphism classes of Steiner triple
systems in a binary 1-perfect code of length 15. The distribution
in Table \ref{tab:spec} is perhaps more even than one might
have guessed.

\begin{table}[htbp]
\begin{center}
\caption{Sizes of sets of Steiner triple systems}
\label{tab:spec}
\begin{tabular}{rrrr}\hline
Size & \# & Size & \#\\ \hline
1 & 437    &8 & 321 \\  
2 & 753	   &9 & 489 \\  
3 & 581	   &10 &110 \\  
4 & 895	   &11 &48  \\	   
5 & 651	   &12 &95  \\	   
6 & 1\,090 &13 &19  \\	    
7 & 452	   &14 &42  \\\hline
\end{tabular}
\end{center}
\end{table}  

If all Steiner triple systems in a code are isomorphic, then the
code is said to be \emph{homogeneous}. By Table \ref{tab:spec}, there
are 437 homogeneous binary
1-perfect codes of length 15. This information is
further refined in Table \ref{tab:hom} by showing which Steiner triple
systems occur in homogeneous codes and in how many they occur.

\begin{table}[htbp]
\begin{center}
\caption{Steiner triple systems in homogeneous codes}
\label{tab:hom}
\begin{tabular}{rrrrrr}\hline
Index & \# & Index & \# & Index & \# \\ \hline
1 & 3  & 9 & 36  & 17 & 10  \\
2 & 23 & 10& 36  & 19 & 9   \\
3 & 15 & 11 & 27 & 22 & 8   \\
4 & 63 & 12 & 7  & 25 & 19  \\
5 & 36 & 13 & 26 & 26 & 19  \\
7 & 5	 & 14 & 18 & 29 & 7 \\
8 & 60 & 16 & 6  & 61 & 4   \\\hline
\end{tabular}
\end{center}
\end{table}

In an analogous way, we may discuss the occurrence of Steiner
quadruple systems in extended binary \mbox{1-perfect codes}. However, since
as many as $15\,590$ (out of a total of $1\,054\,163$)
Steiner quadruple systems of order 16 
occur in extended \mbox{1-perfect codes}, a table analogous to 
Table \ref{tab:sts} would be far too big for this article.
Consequently, we only tabulate, in Table \ref{tab:spec2},
the distribution of the number of isomorphism classes
of Steiner quadruple
systems in extended binary
\mbox{1-perfect} codes of length 16. There are exactly
101 such codes that are homogeneous with respect to Steiner
quadruple systems.

\begin{table}[htbp]
\begin{center}
\caption{Sizes of sets of Steiner quadruple systems}
\label{tab:spec2}
\begin{tabular}{rrrrrr}\hline
Size & \# & Size & \# & Size & \# \\ \hline
1 & 101  & 11 & 91  & 21 & 63 \\
2 & 97   & 12 & 142 & 22 & 28 \\
3 & 77   & 13 & 33  & 23 & 2  \\
4 & 180  & 14 & 109 & 24 & 75 \\
5 & 132  & 15 & 41  & 25 & 4  \\
6 & 172  & 16 & 94  & 28 & 40 \\
7 & 114  & 17 & 38  & 32 & 21 \\
8 & 178  & 18 & 59  & 48 & 2  \\
9 & 93   & 19 & 31  &    &    \\
10 & 131 & 20 & 17  &    &    \\\hline
\end{tabular}
\end{center}
\end{table}  

\section{Structure of $i$-Components}

\label{sect:i}

Consider a binary one-error-correcting code $C$ and a nonempty
subcode $D \subseteq C$. If we get another one-error-correcting code from
$C$ by complementing coordinate $i$ exactly in the words
belonging to $D$, then $D$ is said to be an \mbox{$i$-\emph{component}}
of $C$ and the operation is called \emph{switching}.
An $i$-component is \emph{minimal} if it is not
a superset of a smaller \mbox{$i$-component.} The reader is referred
to \cite{S01} for a more thorough discussion of $i$-components.

Any process that transforms a perfect code into another by changing
values in a single coordinate can be accomplished by switching, because
the codewords that are changed form an $i$-component by
definition. An extension followed by a puncturing 
can be viewed as such a process; hence
all codes that have equal extension can be transformed into each other
by switching.

The \emph{minimum distance graph} of a code consists of one
vertex for each codeword and one edge for each pair
of codewords whose mutual distance equals the minimum
distance of the code. 
All minimal $i$-components can 
be determined by a straightforward algorithm: for
a prescribed value of $i$, construct the minimum distance
graph and remove all edges but those connecting two codewords
that differ in coordinate $i$. The connected 
components of this graph---for 1-perfect codes with length $n \ge 15$
there are at least two of \mbox{them~\cite[Proposition 6]{S01}}---form 
the minimal $i$-components of the code. The minimal $i$-components
partition the code, and any $i$-component is a union of minimal ones.

The distribution of sizes of minimal $i$-components is presented
in Table \ref{tbl:i}. Each row lists the number of sets of
given sizes as well as the number of such partitions (whose
total number is $15\cdot 5\,983 = 89\,745$).
It has been known that partitions
with 2 sets of size 1\,024 as well as 16 sets of size 128 exist,
cf.~\cite{S01}.

\begin{table}[htbp]
\begin{center}
\caption{Sizes of minimal $i$-components}
\label{tbl:i}
\begin{tabular}{rrrrrrr}\hline
128 & 256 & 512 & 768 & 896 & 1\,024 & \#\\ \hline
 16 &     &     &     &     &      &  1\,030  \\
  8 &   4 &     &     &     &      &  1\,536  \\
  8 &     &   2 &     &     &      &  2\,817  \\
  4 &   6 &     &     &     &      &   616  \\
  4 &   2 &   2 &     &     &      &  2\,048  \\
  4 &     &     &   2 &     &      &  2\,587  \\
  2 &     &     &     &   2 &      &   458  \\
    &   8 &     &     &     &      &  1\,023  \\
    &   4 &   2 &     &     &      &  2\,783  \\
    &   2 &     &   2 &     &      &  3\,049  \\
    &     &   4 &     &     &      &  7\,565  \\
    &     &     &     &     &    2 & 64\,233  \\\hline
\end{tabular}
\end{center}
\end{table}

As mentioned above, $i$-components and switching are
means of constructing new codes from old ones. Codes
that can be obtained from each other by a series of 
switches (possibly in different coordinates)
form a \emph{switching class}. (Malyugin \cite{M99,M06} considers 
a more restricted set of transformations that partition 
the switching classes further.)
By \cite{PL99b}, the binary \mbox{1-perfect} codes 
of length 15 are partitioned into at least two switching classes;
we are now able to compute the exact structure of the 
switching classes. 

There are 9 switching classes for the binary 1-perfect codes of length
15, and their sizes are 5\,819, 153, 3, 2, 2, 1, 1, 1, and 1. In 
particular, this gives a method for obtaining codes with (full) rank 15, which have been hard to construct.
The class with 5\,819 codes in fact 
contains all codes with full rank, except two; all codes with rank
11 (the Hamming code), 12, and 13 are also in this class. 
Phelps and LeVan \cite{PL99b} found one of the switching 
classes of size 2.

The two full-rank codes that are not in the switching class of the
Hamming code have one more code in their switching class, a code
with rank 14 (so one may say that all binary 1-perfect codes of
length 15 can be obtained by known constructions). 
These two full-rank codes have kernels of size 2 and
4, and their automorphism groups have orders 336 and 672, respectively.
Both of the codes have an automorphism group which is
the direct product of the
kernel and a group isomorphic to
$\mbox{PSL}(3,2)$, which has order 168; this
group partitions the coordinates into two orbits of size 7 and 
one of size 1.
Indeed, note that $\mbox{PSL}(3,2)$ is the group 
of symmetries of the Hamming code of length 7.

One may generalize the concept of $i$-components to that of
$\alpha$-components; see \cite{AS97}, 
\cite{S08}, and their references. 
An \mbox{$\alpha$-component,} where 
$\alpha \subseteq \{1,2,\ldots ,n\}$, is an $i$-component for
all $i \in \alpha$. We call an $\alpha$-component \emph{trivial}
if it is the full code or if $|\alpha| = 1$.
It turns out that nontrivial $\alpha$-components 
of the binary 1-perfect codes of length 15 
consist of 1\,024 codewords with $|\alpha| \in \{2,3\}$. 

The authors are confident with the double-counting argument 
used in \cite{OP09} for validating the classification of the binary 
\mbox{1-perfect}
codes of length 15; anyway, the fact that no new codes were
encountered in the switching classes further reinforces this
confidence.

\section{Defining Sets of 1-Perfect Codes}

\label{sect:define}

A \emph{defining set} of a combinatorial object is a 
part of the object that uniquely determines the complete
object. The term \emph{unique} should here be interpreted in the 
strongest sense, that is, there should be exactly 
one way of doing this, \emph{not} one way up to isomorphism.

Avgustinovich \cite{A95} gave a brief and elegant proof
(which is repeated in \cite{H08}) of the following result.

\begin{Thm}
\label{thm:def1}
The codewords of weight $(n-1)/2$ (alternatively, weight
$(n+1)/2$) form a defining set of
any binary 1-perfect code of length $n$.
\end{Thm}

Avgustinovich and Vasil'eva
\cite{AV06} were further able to prove the following related
result.

\begin{Thm}
\label{thm:def2}
The codewords of 
weight $w$ with $w \leq (n+1)/2$ form a defining set for the
codewords of weight smaller than $w$ of
any binary 1-perfect code of length $n$.
\end{Thm}

One may ask whether it is possible to strengthen these results
by proving that the codewords of weight $(n-3)/2$ or
any other weight smaller than $(n-1)/2$ form a
defining set for a binary
1-perfect code of length $n$. We shall now prove
that this is not possible in general. In fact, the
theorem will be even stronger than that.

\begin{Thm}
\label{thm:switch}
The Hamming code of length $n$ with $n \geq 7$ 
has no defining set consisting of codewords 
all of whose weights differ from 
$(n-1)/2$ and $(n+1)/2$.
\end{Thm}

\begin{IEEEproof}
The Hamming code of length $n$ has a parity check matrix
\[
\mathbf{H} = 
\left(\begin{array}{ccc}
\mathbf{0} & \mathbf{1} & 1\\
\mathbf{A} & \mathbf{A} & \mathbf{0}\\
\end{array}\right),
\]
where $\mathbf{A}$ is a parity check matrix for the 
Hamming code of length $(n-1)/2$ and
$\mathbf{1}$ is an all-one vector.
It can be easily checked that all words in the set
\[
S = \{(\mathbf{1+x}\ \mathbf{x}\ |\mathbf{x}|): \mathbf{x} 
\in \F_2^{(n-1)/2},\ \mathbf{Ax} = \mathbf{0}\}
\]
are codewords of the Hamming code 
if $n \geq 7$; $|\mathbf{x}|$ is the weight of $\mathbf{x}$ 
modulo $2$.

For an arbitrary word $\mathbf{c} \in S$, consider a word 
$\mathbf{c'}$ in the Hamming code such that
$d(\mathbf{c},\mathbf{c'})=3$ and $\mathbf{c}$ and
$\mathbf{c'}$ differ in the last coordinate. Since
two column vectors of $\mathbf{H}$ that add
to $(1\ \mathbf{0})$ can only have the form
$(0\ \mathbf{a})$ and $(1\ \mathbf{a})$, it follows
that $\mathbf{c'} \in S$. 

Consequently, $S$ is an $i$-component
with respect to the last coordinate, cf.\ Section~\ref{sect:i}. 
A switch in this component produces a different code
with changes made only to the last coordinate. Since
the words of $S$ have weight $(n-1)/2$ considering all but
the last coordinate, the transformed codewords (old and new)
have only weights $(n-1)/2$ and $(n+1)/2$. Hence,
no sets of codewords all weights of which differ from
$(n-1)/2$ and $(n+1)/2$ can form a defining set for 
Hamming codes of length greater than or equal to 7.
\end{IEEEproof}

For specific codes one can find examples of various
defining sets. Examination of the classified codes 
of length 15 shows
that there are cases where the codewords of weight
$4$ form a defining set but no cases where this holds 
for some weight smaller than 4 (this follows from 
Table~\ref{tab:sts} and a consideration of the case
where the smallest weight is 1, for which the words of
weight 3 form a partial Steiner triple system).
The existence of defining sets of weights 5 and 6
follows by using the result for weight 4 and
Theorem~\ref{thm:def2}. The cases of weights greater
than 9 are analogous.

One interesting observation was made in the study of this
property. Namely, there are binary 1-perfect codes of
length 15, whose codewords of weight
7 are a proper subset of the codewords of the same weight in another code. 
In other words, this means that
there are codes whose codewords of weight 7 form a defining
set only under the assumption that this set of words 
contains all codewords of weight 7. This result can be
generalized.

\begin{Thm}
\label{thm:variant}
Theorem \ref{thm:def1} holds for $n \geq 7$ 
only under the assumption that
the given set of codewords of weight $(n-1)/2$ 
(alternatively, weight $(n+1)/2$) is complete.
\end{Thm}

\begin{IEEEproof}
We shall prove that there exist two perfect codes of
length $n$, $C_1$ and $C_2$, so that the codewords of 
weight $(n-1)/2$ of $C_1$ is a proper subset of those
in $C_2$. We let $C_1$ be the Hamming code, defined
by ${\bf H}$ as in the proof of Theorem~\ref{thm:switch}.
Similarly to that proof, we consider $i$-components 
with respect to the last coordinate, but here we focus
on the codewords in 
\[
S = \{(\mathbf{x}\ \mathbf{x}\ |\mathbf{x}|): \mathbf{x} 
\in \F_2^{(n-1)/2},\ \mathbf{Ax} = \mathbf{0}\}.
\] 

Analogously to the argument in
the proof of Theorem~\ref{thm:switch}, $S$ is an $i$-component.
If $wt(\mathbf{x})$ is odd, then the weight of
$(\mathbf{x}\ \mathbf{x}\ |\mathbf{x}|)$ is odd 
(and there are such codewords with weight $(n-1)/2$)
and a switch produces a word with even weight.
If $wt(\mathbf{x})$ is even, say $2v$, then the weight of
$(\mathbf{x}\ \mathbf{x}\ |\mathbf{x}|)$ is $4v$ and
after the switch it becomes $4v+1$, and therefore
cannot equal $(n-1)/2$, which is of the form $2^s-1$,
for $n \geq 7$. Consequently, the codewords of
weight $(n-1)/2$ of the new code $C_2$ obtained
by the switch is a proper subset of the codewords with the
same weight in $C_1$.
\end{IEEEproof}

\section{Systematic 1-Perfect Codes}

\label{sect:syst}

A binary code of size $2^k$ is said to be \emph{systematic}
if there are $k$ coordinates such that the codewords restricted
to these coordinates contain all possible
$k$-tuples; otherwise it is said to be \emph{nonsystematic}. 
It is known \cite{AS96,PL99} that nonsystematic
binary 1-perfect codes exist for all admissible lengths greater
than or equal to 15. It turns out that there are 13 nonsystematic
binary 1-perfect codes of length 15 that extend to 12 nonsystematic codes.

The following invariant is closely related to the concept of
systematic binary codes. The set
\[
\st(C) = \{\supp({\bf x}-{\bf y}) : {\bf x},{\bf y}\in C,\ 
d({\bf x},{\bf y})=3 \}
\]
must obviously have size 
between $\binom{n}{2}/3$ (the size of a Steiner triple
system of order $n$) and $\binom{n}{3}$ when $C$ is a binary
\mbox{1-perfect} code of length $n$; for $n=15$ these bounds are 
$35$ and $455$, respectively.
The distribution of the values of $|\st(C)|$
is shown in Table~\ref{tab:st} for the 1-perfect codes of length 15.

\begin{table}[htbp]
\begin{center}
\caption{Values of $|\st(C)|$}
\label{tab:st}
\begin{tabular}{rr@{\hspace*{2.5mm}}rr@{\hspace*{2.5mm}}rr@{\hspace*{2.5mm}}rr}\hline
$|\st(C)|$ & \# & $|\st(C)|$ & \# & $|\st(C)|$ & \# & $|\st(C)|$ & \# \\ \hline
35 & 1    & 157 & 32    & 212 & 17   & 285 & 34  \\
55 & 1    & 159 & 119   & 213 & 163  & 289 & 1   \\
59 & 2    & 161 & 34    & 214 & 3    & 305 & 3   \\
63 & 15   & 163 & 67    & 215 & 205  & 306 & 2   \\
85 & 3    & 165 & 38    & 216 & 11   & 309 & 2   \\
87 & 1    & 167 & 104   & 217 & 2    & 311 & 8   \\
89 & 2    & 169 & 108   & 218 & 57   & 315 & 1   \\
91 & 3    & 171 & 135   & 219 & 70   & 317 & 1   \\
93 & 3    & 173 & 38    & 220 & 17   & 321 & 3   \\
95 & 7    & 175 & 172   & 221 & 2    & 329 & 3   \\
97 & 4    & 177 & 29    & 222 & 8    & 331 & 1   \\
99 & 30   & 179 & 230   & 224 & 52   & 333 & 2   \\
101 & 47  & 181 & 73    & 225 & 2    & 335 & 4   \\
103 & 49  & 182 & 4     & 229 & 1    & 336 & 4   \\
105 & 31  & 183 & 246   & 231 & 1    & 337 & 3   \\
107 & 184 & 185 & 113   & 233 & 3    & 341 & 2   \\
109 & 91  & 187 & 214   & 237 & 14   & 345 & 11  \\
111 & 76  & 189 & 49    & 239 & 1    & 348 & 1   \\
113 & 97  & 190 & 6     & 241 & 25   & 349 & 1   \\
115 & 50  & 191 & 473   & 243 & 5    & 353 & 11  \\
117 & 22  & 193 & 284   & 245 & 44   & 357 & 2   \\
119 & 4   & 194 & 4     & 247 & 1    & 361 & 17  \\
127 & 1   & 195 & 221   & 249 & 1    & 365 & 7   \\
129 & 2   & 197 & 95    & 253 & 3    & 366 & 2   \\
131 & 1   & 198 & 3     & 255 & 2    & 369 & 11  \\
133 & 1   & 199 & 200   & 257 & 4    & 373 & 11  \\
135 & 3   & 200 & 4     & 261 & 5    & 375 & 7   \\
137 & 6   & 201 & 236   & 263 & 5    & 377 & 17  \\
139 & 4   & 202 & 3     & 265 & 2    & 388 & 1   \\
141 & 1   & 203 & 120   & 269 & 16   & 404 & 1   \\
143 & 6   & 205 & 77    & 271 & 4    & 414 & 2   \\
145 & 6   & 206 & 5     & 273 & 7    & 427 & 6   \\
147 & 13  & 207 & 151   & 275 & 6    & 438 & 2   \\
149 & 5   & 208 & 1     & 277 & 16   & 455 & 6   \\
151 & 10  & 209 & 181   & 279 & 6    &     &     \\
153 & 18  & 210 & 4     & 281 & 17   &     &     \\
155 & 43  & 211 & 271   & 283 & 8    &     &     \\\hline
\end{tabular}
\end{center}
\end{table}  

Generalizing the concept of independent sets in graphs,
a subset $S$ of the vertices of a hypergraph is said to be 
\emph{independent} if none of the edges is included in $S$.
Viewing $\st(C)$ of a code $C$ of length $n = 2^m-1$ as a 
3-uniform hypergraph, if the independence number of this 
graph---denoted by $\alpha(C)$---is
smaller than $m$, then $C$ in nonsystematic \cite{AS96}.
In particular, if $|\st(C)| = \binom{n}{3}$, then 
$\alpha(C)=2$ and the code is nonsystematic.

Out of the 13 nonsystematic 1-perfect codes of length 15, 
six indeed have $|\st(C)| = \binom{15}{3} = 455$. Out of the others,
six have $|\st(C)| = 427$ with $\alpha(C)= 3$, and one has 
$|\st(C)| = 231$ with $\alpha(C)=8$. The distribution
of the independence numbers of all codes, systematic as well as 
nonsystematic, is presented in Table~\ref{tab:ind}.

\begin{table}[htbp]
\begin{center}
\caption{Independence numbers of codes}
\label{tab:ind}
\begin{tabular}{rrrr}\hline
$\alpha(C)$ & \# & $\alpha(C)$ & \# \\\hline
2 & 6  & 5 & 107 \\
3 & 6  & 6 & 238 \\
4 & 41 & 8 & 5\,585 \\\hline
\end{tabular}
\end{center}
\end{table}

Examples of nonsystematic binary 1-perfect codes of length 15
with $|\st(C)|=455$ and $|\st(C)|=427$ were earlier 
obtained in \cite{PL99}, and 12 inequivalent nonsystematic 
codes were encountered in
\cite{M06}; see also \cite{R97}.
The fact that there is a nonsystematic code $C$ of
length $2^m-1$ such that $\alpha(C)>m$ shows that
the above mentioned sufficient condition for 
a \mbox{1-perfect} code $C$ to be nonsystematic is not necessary,
a question asked in \cite{PL99}. 

For a binary 1-perfect code of length 15 with $\alpha(C)=8$, 
let $S$ be the complement
of a maximum independent set. A counting argument shows that
each STS(15) of the code contains 7 triples on $S$, and hence each STS(15)
contains an STS(7) on S. Moreover, for the code with $|\st(C)|=231$,
$\st(C)$ contains exactly those 3-subsets that intersect $S$ in
1 or 3 coordinates.
Also the 3-subsets for the codes 
with $|\st(C)|=427$ have a combinatorial explanation: the missing
3-subsets form an STS(15) with one point and the blocks intersecting
that point removed, in other words, a 3-GDD of type $2^7$.

\section{Embedding One-Error-Correcting Codes}

\label{sect:embed} 

Avgustinovich and Krotov \cite{AK09} show that any binary one-error-correcting
code of length $m$ can be embedded (after appending a
zero vector of appropriate length to the codewords) into a binary
1-perfect code of length $2^m-1$. One may further ask for the
shortest 1-perfect code into which such a code can be embedded.
For example, any binary one-error-correcting code of length $4$---there are 
three inequivalent such codes, \{0000\}, $\{0000,1110\}$, and 
$\{0000,1111\}$---can be embedded into the 
(unique) 1-perfect code of length 7, but this does not hold for
all codes of length 5 (because the Hamming code of length 7 does not 
contain a pair of codewords with mutual distance 5).

The occurrence of codes of length greater than or equal to 5 in
the classified codes was checked. It turns out that for lengths
5 and 6, all inequivalent one-error-correcting
codes can be found
in a binary 1-perfect code of length 15, but not all such codes of
length 7 can be found. Out of several examples, here is one
code (of size 10) that is not embeddable in a
binary 1-perfect code of length 15:
\[
\begin{tabular}{lllll}
0000000 & 
0001111	& 
0101100	& 
0110110	& 
0111011	\\ 
1001001 &
1010111 &
1011100 &
1101010 &
1110001.\\ 
\end{tabular}
\]

One could further consider the stricter requirement that
a code should be embedded in a perfect code in such a
way that it is not a subset of
another embedded code. The construction in \cite{AK09}
indeed gives codes with this strong property.
This requirement is rather
restrictive as, for example, the code $\{0000\}$ does
not fulfill it with respect to the Hamming code of length 7.

The largest embedded codes with the property of
not being subcodes of other embedded codes are directly
related to certain fundamental properties. For example,
a binary 1-perfect code of length $2^m-1$ is systematic if and only if 
there exists an $m$-subset of coordinates for which the 
maximum size is 1. In other words, minimizing the maximum size
over all $m$-subsets of coordinates should result in size 1.
Instead maximizing the maximum size over subsets
leads to the concept of \emph{cardinality-length profile} (CLP), 
from which one may obtain the \emph{generalized Hamming
weight hierarchy} of a code; see \cite{EV98,LV95}.

For a code $C \subseteq \F_2^n$, the
cardinality-length profile $\kappa_i(C)$, $1 \leq i \leq n$, 
is defined as
\[
\kappa_i(C) = \max_D \log_2 |D|,
\]
where $D \subseteq C$ and all words in $D$ must coincide in 
$n-i$ coordinates. The profiles of the binary 1-perfect
codes of length 15 are listed in Table~\ref{tbl:clp} in the
nonlogarithmic 
form ${\kappa'_1(C)},{\kappa'_2(C)},\ldots$, where
$\kappa'_i(C)=2^{\kappa_i(C)}$, together
with the number of codes with such profiles.

\begin{table*}[htbp]
\begin{center}
\caption{Cardinality-Length Profiles of Codes}
\label{tbl:clp}
\begin{tabular}{r@{\hspace*{3mm}}r@{\hspace*{3mm}}r@{\hspace*{3mm}}r@{\hspace*{3mm}}r@{\hspace*{3mm}}r@{\hspace*{3mm}}r@{\hspace*{3mm}}r@{\hspace*{3mm}}r@{\hspace*{3mm}}r@{\hspace*{3mm}}r@{\hspace*{3mm}}r@{\hspace*{3mm}}r@{\hspace*{3mm}}r@{\hspace*{3mm}}rr}\hline
$\kappa'_1(C)$ &$\kappa'_2(C)$ &$\kappa'_3(C)$ &$\kappa'_4(C)$ &
$\kappa'_5(C)$ &$\kappa'_6(C)$ &$\kappa'_7(C)$ &$\kappa'_8(C)$ &
$\kappa'_9(C)$ &$\kappa'_{10}(C)$ &$\kappa'_{11}(C)$ &
$\kappa'_{12}(C)$ &$\kappa'_{13}(C)$&$\kappa'_{14}(C)$&
$\kappa'_{15}(C)$ & \#\\\hline
1&1&2&2&4&6&10&16&32&64&128&256&512&1\,024&2\,048&10\\
1&1&2&2&4&7&10&16&32&64&128&256&512&1\,024&2\,048&1\\
1&1&2&2&4&7&11&16&32&64&128&256&512&1\,024&2\,048&23\\
1&1&2&2&4&8&11&16&32&64&128&256&512&1\,024&2\,048&3\\ 
1&1&2&2&4&6&12&16&32&64&128&256&512&1\,024&2\,048&7\\
1&1&2&2&4&7&12&16&32&64&128&256&512&1\,024&2\,048&26\\
1&1&2&2&4&8&12&16&32&64&128&256&512&1\,024&2\,048&219\\
1&1&2&2&4&7&13&16&32&64&128&256&512&1\,024&2\,048&7\\
1&1&2&2&4&8&13&16&32&64&128&256&512&1\,024&2\,048&48\\
1&1&2&2&4&7&14&16&32&64&128&256&512&1\,024&2\,048&6\\ 
1&1&2&2&4&8&14&16&32&64&128&256&512&1\,024&2\,048&34\\
1&1&2&2&4&8&15&16&32&64&128&256&512&1\,024&2\,048&14\\
1&1&2&2&4&8&16&16&32&64&128&256&512&1\,024&2\,048&5585\\\hline
\end{tabular}
\end{center}
\end{table*}

The fact that the number of codes in the
last row of Table~\ref{tbl:clp} equals the number of 
binary 1-perfect 
codes of length 15 that do not have full rank is in accordance
with \mbox{\cite[Corollary 4.7]{EV98}}. By 
\cite[Proposition 4.5]{EV98} we already know that
$8 < \kappa'_7(C) < 16$; the new results reveal that
$\kappa'_7(C)$ can attain every value in this 
interval except 9. The only other value of $i$ for
which $\kappa'_i(C)$ may be different for codes that have
and do not have full rank is $i=6$; however, whereas 
$\kappa'_6(C) \in \{6,7\}$ is only possible for full-rank codes, 
$\kappa'_6(C) = 8$ is possible for both types.

Also the cardinality-length profiles of the extended 
binary \mbox{1-perfect} codes of length 16 are listed
in nonlogarithmic form, in Table~\ref{tbl:clp2}. 

\begin{table*}[htbp]
\begin{center}
\caption{Cardinality-Length Profiles of Extended Codes}
\label{tbl:clp2}
\begin{tabular}{r@{\hspace*{2.1mm}}r@{\hspace*{2.1mm}}r@{\hspace*{2.1mm}}r@{\hspace*{2.1mm}}r@{\hspace*{2.1mm}}r@{\hspace*{2.1mm}}r@{\hspace*{2.1mm}}r@{\hspace*{2.1mm}}r@{\hspace*{2.1mm}}r@{\hspace*{2.1mm}}r@{\hspace*{2.1mm}}r@{\hspace*{2.1mm}}r@{\hspace*{2.1mm}}r@{\hspace*{2.1mm}}r@{\hspace*{2.1mm}}rr}\hline
$\kappa'_1(C)$&$\kappa'_2(C)$ &$\kappa'_3(C)$ &$\kappa'_4(C)$ &
$\kappa'_5(C)$ &$\kappa'_6(C)$ &$\kappa'_7(C)$ &$\kappa'_8(C)$ &
$\kappa'_9(C)$ &$\kappa'_{10}(C)$ &$\kappa'_{11}(C)$ &
$\kappa'_{12}(C)$ &$\kappa'_{13}(C)$&$\kappa'_{14}(C)$&
$\kappa'_{15}(C)$ & $\kappa'_{16}(C)$ & \#\\\hline
1&1&1&2&2&4&6&10&16&32&64&128&256&512&1\,024&2\,048&   6\\ 	    
1&1&1&2&2&4&7&10&16&32&64&128&256&512&1\,024&2\,048&   1\\ 	    
1&1&1&2&2&4&7&11&16&32&64&128&256&512&1\,024&2\,048&  13\\ 	    
1&1&1&2&2&4&8&11&16&32&64&128&256&512&1\,024&2\,048&   2\\ 	    
1&1&1&2&2&4&6&12&16&32&64&128&256&512&1\,024&2\,048&   5\\ 	    
1&1&1&2&2&4&7&12&16&32&64&128&256&512&1\,024&2\,048&  13\\ 	    
1&1&1&2&2&4&8&12&16&32&64&128&256&512&1\,024&2\,048&  69\\ 	    
1&1&1&2&2&4&7&13&16&32&64&128&256&512&1\,024&2\,048&   6\\ 	    
1&1&1&2&2&4&8&13&16&32&64&128&256&512&1\,024&2\,048&  20\\ 	    
1&1&1&2&2&4&7&14&16&32&64&128&256&512&1\,024&2\,048&   6\\ 	    
1&1&1&2&2&4&8&14&16&32&64&128&256&512&1\,024&2\,048&  21\\ 	    
1&1&1&2&2&4&8&15&16&32&64&128&256&512&1\,024&2\,048&  13\\ 	    
1&1&1&2&2&4&8&16&16&32&64&128&256&512&1\,024&2\,048&1990\\\hline
\end{tabular}
\end{center}
\end{table*}

One more concept can be related to the discussion in this
section. An $\mbox{OA}_\lambda (t,k,q)$ 
\emph{orthogonal array} of index $\lambda$, strength $t$,
degree $k$, and order $q$ is a $k \times N$ array
with entries from $\{0,1,\ldots ,q-1\}$ and the property that 
every $t \times 1$ column vector appears exactly $\lambda$ times
in every $t \times N$ subarray; necessarily $N = \lambda q^t$.

Binary 1-perfect codes of length 15 can be
viewed as $15 \times 2\,048$ 
arrays and their extensions as $16 \times 2\,048$ arrays.
It is then obvious from Tables~\ref{tbl:clp} and
\ref{tbl:clp2} that these are
$\mbox{OA}_{16}(7,15,2)$ and
$\mbox{OA}_{16}(7,16,2)$ orthogonal arrays, respectively.
In fact, we shall prove (a general result) implying that
there are no other orthogonal arrays with
these parameters. The following result by Delsarte~\cite{D73}
is of central importance in the main proof.

\begin{Thm}
\label{thm:delsarte}
An array is an orthogonal array of strength $t$
if and only if the MacWilliams transform of the distance distribution
of the code formed by the columns of the array has entries
$A'_0 = 1$, $A'_1 = A'_2 = \cdots = A'_t = 0$.
\end{Thm}

For information on the MacWilliams transform in
general and the application to orthogonal arrays in particular, see
\mbox{\cite[Ch.~5]{MS77}} and \cite[Ch.~4]{HSS99}, respectively.
With standard techniques---frequently used in a similar
context, see, for example, \cite{N79,N86,X05}---one can now 
prove the following result; the part of the
proof showing that these codes are orthogonal arrays
can be found in
several places, including \cite[Theorem 4.4]{EV98}.

\begin{Thm}
\label{thm:oa}
Every $\mbox{OA}_{\lambda}(t,n,2)$ with $n=2^m-1$, \mbox{$t=(n-1)/2$}, and
$\lambda = 2^{2^{m-1}-m}$ corresponds to a 1-perfect binary code
of length $n$, and vice versa.
Every $\mbox{OA}_{\lambda}(t,n,2)$ with $n=2^m$ and $t$ and $\lambda$ as earlier
corresponds to an extended 
1-perfect binary code of length $n$, and vice versa.
\end{Thm}
\begin{IEEEproof}
Perfect codes and their extensions have a unique
distance distribution. Therefore, the minimum distance
of the dual of the (extended) Hamming code of length $n$
gives the (maximum possible) value of $t$ in 
Theorem~\ref{thm:delsarte} for any (extended) 1-perfect
code. A simplex code with a minimum distance of
$(n+1)/2$ is the dual of a Hamming code, and a 
first order Reed-Muller code with a minimum distance
of $n/2$ is the dual of an extended Hamming code. This
shows that every (extended) 1-perfect code is an orthogonal
array with the given parameters. 

Going in the opposite direction, it is clear that
the code obtained from the orthogonal array
$\mbox{OA}_{\lambda}(t,n,2)$ with $n=2^m-1$, 
\mbox{$t=(n-1)/2$}, and
$\lambda = 2^{2^{m-1}-m}$
has the length and cardinality of a \mbox{1-perfect} code.
Let $A'_i$ be the MacWilliams transform of the distance distribution
of the code. By Theorem~\ref{thm:delsarte} we have $A'_0 = 1$,
$A'_1 = \cdots = A'_t = 0$. Moreover, 
\[
\sum_{i=0}^n A'_i = \frac{2^n}{\lambda 2^t} = 
\frac{2^{2^m-1}}{2^{2^{m-1}-m}2^{((2^m-1)-1)/2}} = 
n+1.
\]
By taking the MacWilliams transform of $A'_i$
we get the distance distribution $A_i$.
In particular,
\[
(n+1)A_1 = A'_0 P_1(0) + A'_{t+1}P_1(t+1) + \cdots + A'_n P_1(n),
\]
where $P_j(i)$ is the Krawtchouk polynomial.
We have $P_1(i) = n - 2i$ and thus
\begin{eqnarray*}
(n+1)A_1 & = & n  -A'_{t+1} -3A'_{t+2} - \cdots -nA'_n\\
         & \le & n -(A'_{t+1} + A'_{t+2} + \cdots + A'_n)\\
         & = & 0.
\end{eqnarray*}
Since $A_1 \geq 0$, we must have 
$n  -A'_{t+1} -3A'_{t+2} - \cdots -nA'_n = 0$, which
together with $\sum_{i=t+1}^n A_i = n$ has only
one possible solution: 
\[
A'_{t+1} = n,\
A'_{t+2} = A'_{t+3} = \cdots = A'_n = 0.
\]
This solution $A'_i$ is the MacWilliams
transform of the weight distribution of the Hamming code,
so $A_i$ is the distance distribution of the Hamming code.
Thereby the code has minimum distance $3$ and is 1-perfect.

Finally, we have to show that every $\mbox{OA}_{\lambda}(t,n,2)$ 
with $n=2^m$ and $t$ and $\lambda$ as above---which
gives a code with the same length and size as 
an extended 1-perfect binary code---leads to a code with 
minimum distance 4. This follows from the previous part of
the proof. Namely, deleting any row of the orthogonal array
$\mbox{OA}_{\lambda}(t,n,2)$ gives an 
$\mbox{OA}_{\lambda}(t,n-1,2)$, which we know that
corresponds to a code with minimum distance 3. If 
the $\mbox{OA}_{\lambda}(t,n,2)$ would correspond to a code
with minimum distance 3 (or less), then there would be two 
columns at mutual distance 3 (or less), and a row of 
the orthogonal array could be deleted so that the mutual 
distance between the columns would become 2 (or less). This
contradiction completes the proof.
\end{IEEEproof}

\begin{Cor}
\label{cor:oa}
The number of isomorphism classes of 
$\mbox{OA}_{16}(7,15,2)$ orthogonal arrays 
is 5\,983, and 
the number of isomorphism classes of 
$\mbox{OA}_{16}(7,16,2)$ orthogonal arrays 
is 2\,165.
\end{Cor}

Classification results for some other related 
orthogonal arrays can also be obtained.

\begin{Thm}
\label{thm:oa2}
Every $\mbox{OA}_{\lambda}(t,n,2)$ with $n=2^m-2$, \mbox{$t=(n-3)/2$} and
$\lambda = 2^{2^{m-1}-m}$ can be obtained by shortening a
1-perfect codes, and 
every $\mbox{OA}_{\lambda}(t,n,2)$ with $n=2^m-1$ and
$t$ and $\lambda$ as earlier
can be obtained by shortening an
extended 1-perfect code.
\end{Thm}
\begin{IEEEproof}
By \cite[Theorem 2.24]{HSS99}, an $\mbox{OA}_{\lambda}(2t,k,q)$
can be obtained from an $\mbox{OA}_{\lambda}(2t+1,k+1,q)$ 
and any $\mbox{OA}_{\lambda}(2t,k,q)$ can be used to construct
an $\mbox{OA}_{\lambda}(2t+1,k+1,q)$. Application of 
Theorem \ref{thm:oa} completes the proof.
\end{IEEEproof}

We now get two specific results using values calculated
in \cite{OP09}.

\begin{Cor}
\label{cor:oa2}
The number of isomorphism classes of
$\mbox{OA}_{16}(6,14,2)$ orthogonal arrays 
is $38\,408$, and 
the number of isomorphism classes of
$\mbox{OA}_{16}(6,15,2)$ orthogonal arrays 
is $5\,983$.
\end{Cor}

\section{Mixed Perfect Codes}

\label{sect:mixed}

The discussion in Section \ref{sect:syst} focuses on (the structure of)
pairs of codewords with mutual distance 3. For a particular
such structure, which we shall now discuss, one is able to construct
perfect codes with both quaternary and binary coordinates. Moreover,
since this construction is reversible, we obtain a complete
classification of these codes. 

Assume that there exist three coordinates of a binary \mbox{1-perfect} code 
of length 15 such that all codewords can be partitioned into pairs
of words that differ only in those coordinates. In other words,
the kernel of the code has an element with 1s in exactly these
three coordinates. The values of the
pairs in the three coordinates are then $\{000,111\}$,
$\{001,110\}$, $\{010,101\}$, and $\{100,011\}$. It is
not difficult to verify that replacing each original
pair of codewords with one codeword and the three coordinates by
an element from the finite field $\F_4$ (or any alphabet of size 4)
gives a mixed \mbox{1-perfect} codes. This transformation is reversible,
and has been used several times to construct good binary codes 
(of various types, both covering and packing) from codes with 
quaternary coordinates \cite{H88,KP88,O91}.

Moreover, for any set of $t$ elements in the kernel with
weight 3 and disjoint supports, we can obtain a mixed 1-perfect
code with $t$ quaternary coordinates.
By determining all possible mixed 1-perfect codes
that can be obtained in this manner and carrying
out isomorph rejection among these, we found that the number 
of inequivalent 1-perfect
codes over $\F_4^1\F_2^{12}$, $\F_4^2\F_2^{9}$, $\F_4^3\F_2^{6}$, 
$\F_4^4\F_2^{3}$, and $\F_4^5$ are 6\,483, 39, 4, 1, and 1,
respectively. Uniqueness of the quaternary 1-perfect code of length 5
has earlier been proved in \cite{A06}. The orders of the
automorphism groups of
these codes are listed in Tables \ref{tbl:perf1} to \ref{tbl:perf5}.
(The existence of the codes is well known, for example,
via the existence of the quaternary Hamming code of length 5 and 
the construction discussed above.)

\begin{table}[htbp]
\begin{center}
\caption{Automorphism groups of 1-perfect codes over $\F_4^1\F_2^{12}$}
\label{tbl:perf1}
\begin{tabular}{rrrrrr}\hline
$|\Aut(C)|$ & \# & $|\Aut(C)|$ & \# & $|\Aut(C)|$ & \# \\ \hline
8   & 1	    & 768    &  11 &	24\,576 & 8\\   
16  & 12    & 1\,024 & 609 &	32\,768 & 7\\ 
32  & 289   & 1\,536 & 22  &	49\,152 & 3\\ 
64  & 1\,125& 2\,048 & 343 &	65\,536 & 2\\ 
96  & 1	    & 3\,072 & 25  &	98\,304 & 5\\ 
128 & 1\,447& 4\,096 & 154 &	196\,608&  1\\  
192 & 14    & 6\,144 & 12  &	294\,912&  1\\  
256 & 1\,390& 8\,192 & 64  &	589\,824&  1\\  
384 & 14    & 12\,288&  4  &&\\
512 & 892   & 16\,384&  26 &&\\\hline		
\end{tabular}
\end{center}
\end{table}  

\begin{table}[htbp]
\begin{center}
\caption{Automorphism groups of 1-perfect codes over $\F_4^2\F_2^{9}$}
\label{tbl:perf2}
\begin{tabular}{rrrr}\hline
$|\Aut(C)|$ & \# & $|\Aut(C)|$ & \# \\ \hline
128    & 3  &  2\,048 & 7 \\
256    & 6  &  4\,096 & 3 \\	  
512    & 10 &  6\,144 & 1 \\	  
768    & 1  & 36\,864 & 1 \\
1\,024 & 7  & & \\\hline
\end{tabular}
\end{center}
\end{table}  

\begin{table}[htbp]
\begin{center}
\caption{Automorphism groups of 1-perfect codes over $\F_4^3\F_2^{6}$}
\label{tbl:perf3}
\begin{tabular}{rr}\hline
$|\Aut(C)|$ & \#\\ \hline
1\,024 & 2\\
3\,072 & 1\\
9\,216 & 1\\\hline
\end{tabular}
\end{center}
\end{table}  

\begin{table}[htbp]
\begin{center}
\caption{Automorphism groups of 1-perfect codes over $\F_4^4\F_2^{3}$}
\label{tbl:perf4}
\begin{tabular}{rr}\hline
$|\Aut(C)|$ & \#\\ \hline
9\,216        &  1\\\hline
\end{tabular}
\end{center}
\end{table}  

\begin{table}[htbp]
\begin{center}
\caption{Automorphism groups of 1-perfect codes over $\F_4^5$}
\label{tbl:perf5}
\begin{tabular}{rr}\hline
$|\Aut(C)|$ & \#\\ \hline
23\,040       & 1\\\hline
\end{tabular}
\end{center}
\end{table}  

The four code pairs of length 3 listed earlier are in fact cosets of 
the binary Hamming code of length 3, and the outlined
construction is a special case of a general construction \cite{KP88}
that transforms coordinates over $\F_{2^m}$ into cosets of the
Hamming code of length $2^m-1$.

To transform a binary 1-perfect code of length 15 into a \mbox{1-perfect}
code over $\F_8^1\F_2^8$, we may search for a partition of the code 
into 128 subcodes of size 16 with all words in a subcode coinciding 
in 8 given coordinates. However, it turns out that this case can
be proved in a direct way.

\begin{Thm}
\label{thm:f8}
There are exactly 10 inequivalent 1-perfect codes over
$\F_8^1\F_2^8$.
\end{Thm}

\begin{IEEEproof}
Consider a 1-perfect code $C$ over $\F_8^1\F_2^8$; such
a code has size 128. Puncturing $C$ in the 8-ary coordinate
gives a binary code $C'$ of length 8 and minimum distance at
least $3-1=2$. The code $C'$ is unique, and
consists of either all words of even weight or all words
of odd weight. 

Next consider the 8 subcodes $C_i$ obtained by shortening
in the 8-ary coordinate and taking all words whose 8-ary
value is $i$. The codes $C_i$ have length 8 and minimum 
distance 4 (at least 3, but words in $C'$ do not have
odd mutual distances), and since the maximum number
of codewords in a code with these parameters is 16, all codes $C_i$ must
have size 16 ($16\cdot 8=128$).

Consequently, the set of subcodes $C_i$ form an extension of
a partition of $\F_2^7$ into binary 1-perfect codes of
length 7. There are 10 inequivalent such extended 
partitions \cite{P00} and accordingly equally many 
inequivalent 1-perfect codes over $\F_8^1\F_2^8$.
\end{IEEEproof}

The existence of 1-perfect codes over $\F_8^1\F_2^8$ has been
known and follows, for example, from \mbox{\cite[Theorem~2]{HS71}.}
The automorphism groups of these codes can be obtained from 
\cite[Appendix]{P00} and are shown in Table \ref{tbl:perf8}.

\begin{table}[htbp]
\begin{center}
\caption{Automorphism groups of 1-perfect codes over $\F_8^1\F_2^{8}$}
\label{tbl:perf8}
\begin{tabular}{rrrr}\hline
$|\Aut(C)|$ & \# &  $|\Aut(C)|$ & \#\\ \hline
  768   &  1 & 6\,144  &  1\\   
1\,024  &  3 & 8\,192  &  1\\   
2\,688  &  1 & 12\,288 &  1\\   
3\,072  &  1 & 172\,032&  1\\\hline
\end{tabular}
\end{center}
\end{table}

No other 1-perfect codes over $\F_{2^{i_1}}^1\F_{2^{i_2}}^1\cdots
\F_{2^{i_n}}^1$ can be obtained from the binary 1-perfect codes of
length 15, since any such code must have $i_j + i_k \leq 4$ for all
$1\leq j<k \leq n$ \cite[Lemma 1]{L75}.

\section{Various Other Properties}

\label{sect:other}

The current study focuses on properties of general interest;
various other questions in the literature
that can be addressed via the classified
codes include those in \cite[Sect.\ 8]{DD05}.
We conclude the paper by discussing 
a few sporadic results and open problems.

Several (nontrivial) properties of perfect codes
have earlier been proved analytically; we shall here briefly
mention one such property.
The minimum distance graph of a
binary \mbox{1-perfect} code of length 15 is a 35-regular
graph of order 2\,048.
Phelps and LeVan \cite{PL99b} ask whether inequivalent
binary \mbox{1-perfect} codes always have nonisomorphic minimum 
distance graphs. This question is answered in the affirmative
by Avgustinovich in \cite{A01}, building on earlier work by
Avgustinovich and others \cite{A94,SAHH98}. An analogous result
is obtained for extended binary 1-perfect codes in
\cite{MOPS09}, where it is also shown that the
automorphism group of an (extended) binary 1-perfect code
is isomorphic to the automorphism group of its
minimum distance graph for lengths $n \geq 15$.

Let $(n,M,d)$ denote a binary code of length $n$, size $M$,
and minimum distance $d$; such a code with the largest possible 
value of $M$ with the other parameters fixed
is called \emph{optimal}. By shortening binary
1-perfect codes of length 15 up to $i=3$ times we get optimal
$(15-i,2^{11-i},3)$ codes. But do we get all such codes, 
up to equivalence, in this manner? For $i=1$ we do, as shown 
in \cite{B99}; this result was used in \cite{OP09} to classify 
the optimal $(14,1\,024,3)$ codes. But for $i=2$ we
do not, as shown in \cite{OP09b}. 
The general problem is, however, still open.

Two of the open 
problems stated in \cite{EV98} still seem
out of reach even for the case of 
binary 1-perfect codes of length 15.

The \emph{intersection number} of two codes, $C_1$ and $C_2$,
is $|C_1 \cap C_2|$. The intersection number problem asks for
the set of possible intersection numbers of distinct binary
1-perfect codes. Since binary 1-perfect codes are self-complementary,
these intersection numbers are necessarily even. Among other things
it is known that for binary 1-perfect codes of length 15,
0 (trivial) and 2 (by \cite[Theorem 3.2]{EV98})
are intersection numbers and in \cite[Sect.\ III]{EV94} it is
proved that the largest number is $2^{11}-2^7 = 1\,920$. Several
other intersection numbers are known \cite{AHS06}, 
but determining all possible intersection numbers 
seems challenging.

The proof of Theorem \ref{thm:f8} relies on a classification
\cite{P00} of partitions of $\F_2^7$ into binary 1-perfect codes.
This classification problem may be considered for $\F_2^{15}$
as well,
but even the restricted version (stated in \cite{EV98}) with  
16 equivalent codes in the partition seems hopeless.

\section*{Acknowledgments}

The authors thank Olof Heden, Denis Krotov,
Sergey Mal\-yu\-gin,
Ivan Mogilnykh, Faina Solov'eva, Dmitrii Zinov'ev, 
Victor Zinov'ev, and the anonymous reviewers for their 
helpful comments. The proofs of Theorems~\ref{thm:switch} 
and \ref{thm:variant} are based on comments by
one of the referees.

\begin{IEEEbiographynophoto}{Patric R. J. \"Osterg{\aa}rd}
  was born in Vaasa, Finland, in 1965. He received the M.Sc.\ (Tech.)\
  degree in electrical engineering and the D.Sc.\ (Tech.)\ degree in
  computer science and engineering, in 1990 and 1993, respectively,
  both from Helsinki University of Technology TKK, Espoo, Finland.

  From 1989 to 2001, he was with the Department of Computer Science
  and Engineering at TKK. During 1995--1996, he visited Eindhoven
  University of Technology, Eindhoven, The Netherlands.  Since 2000,
  he has been a Professor at TKK---which merged with two other universities into the Aalto University in January 2010---currently in the Department of
  Communications and Networking. He was the Head of the Communications
  Laboratory at TKK in 2006--2007.  He is co-author of
  \emph{Classification Algorithms for Codes and Designs} (Berlin:
  Springer-Verlag, 2006), and, since 2006, co-Editor-in-Chief of the
  \emph{Journal of Combinatorial Designs}.  His research interests
  include algorithms, coding theory, combinatorics, design theory, and
  optimization.

  Dr.\ \"Osterg{\aa}rd is a Fellow of the Institute of Combinatorics
  and its Applications. He is a recipient of the 1996 Kirkman Medal.
\end{IEEEbiographynophoto}

\begin{IEEEbiographynophoto}{Olli Pottonen}
  was born in Helsinki, Finland, in 1984.  He received the M.Sc.\
  (Tech.)\ degree in engineering physics and D.Sc\ (Tech.)\ degree in
  information theory from Helsinki University of Technology TKK,
  Espoo, Finland, in 2005 and 2009, respectively.  He is currently a
  Researcher at the Finnish Defence Forces Technical Research Centre. His
  research interests include discrete mathematics, combinatorics, and
  coding theory.
\end{IEEEbiographynophoto}

\begin{biographynophoto}{Kevin Phelps} was born in New York, New York
(U.S.A.) in May 1948. He received his Bachelor's degree in
Mathematics from Brown University in 1970 and his PhD degree in
Mathematics from Auburn University in 1976. In 1976, he joined
the faculty of Georgia Institute of Technology, School of
Mathematics as an Assistant Professor, and was promoted to
Associate Professor in 1983. In 1987, he joined the faculty at
Auburn University, Division of Mathematics as a Professor. He
became Head of the Department of Discrete and Statistical Sciences
at Auburn University in 1992 and remained in that position until
2004. He has been a Visiting Professor at University of Waterloo,
Canada, at the University of Bielefeld, Germany,  and at the Centre
de Recerca Matem\`{a}tica, Universitat Aut\`{o}noma  de Barcelona,
Spain. His research interests include coding theory, combinatorial
designs and set systems, and cryptography as well as associated
algorithmic and computational problems.
\end{biographynophoto}

\end{document}